\documentclass[aps,twocolumn,showpacs]{revtex4}
\usepackage[dvips]{graphicx}

\begin{document}

\title{Electronic transmission in bent quantum wires} 

\author{Arunava Chakrabarti}

\affiliation{Department of Physics, University of Kalyani, Kalyani,  
West Bengal-741 235, India.} 

\begin{abstract}
Electronic transmission in bent quantum wires modeled by 
the tight binding Hamiltonian, and clamped between ideal, 
semi-infinite leads is studied. The effect of `bending' the chain is 
simulated by introducing a non-zero hopping between the extremities 
of the wire. It is seen that the proximity of the two ends gives rise
to Fano line shapes in the transmission spectrum. Transmission 
properties for both an ordered lattice and a Fibonacci quantum wire 
are discussed. In the quasi-periodic Fibonacci chain, the proximity 
of the two ends of the chain closes all the gaps in the spectrum and the 
spectrum loses its Cantor set character.  

\end{abstract}


\maketitle 

\noindent
\section{Introduction}

Winding chains as models of conducting polymers present several 
interesting features in their electronic spectrum and transport properties
~\cite{evan1,evan2}. 
As a result of winding at one or more places in the chain an electron gets 
an alternative path to hop to a site, which might have been a 
distant neighbor otherwise.
This {\it short cut} path provided to the propagating 
electron leads to non-trivial transport characteristics, sometimes 
leading to groups of states with very large localization length, indicative 
of a metal-insulator transition~\cite{evan1}. The features persist and become  
even richer when more than one chains couple in the transverse direction 
with random winding structure~\cite{evan2}. It has recently been argued that 
such simple models of the conducting polymers where an electron travels 
along a randomly tangled chain has the geometry of a ``small-world network" 
~\cite{quint}, and may lead to a multifractal set of single particle states.

There is another aspect of interest hidden in such simple models that are 
quite relevant in the context of the present development of  mesoscopic physics or of 
nano-technology.
It is now possible to build up `atomic chains' using the scanning tunneling microscope (STM) 
tip as tweezers. The tailor made quantum devices in recent years  
have inspired immense research on quantum wires (QW) and quantum dots (QD) that have been 
major areas in nano-electronics and nano structure physics~\cite{supriyo,ferry}. Electronic 
transport in an array of QD's or in a QW can now be studied in a controllable way
~\cite{gold,cron}, and a wide variety of theoretical studies based on simple orbital 
~\cite{ore}-\cite{mard} tight binding models have been reported in the literature 
that reproduce several basic features of 
quantum transport in QD arrays, QW or molecular wires~\cite{papa}.
\begin{figure}[ht]
{\centering\resizebox*{5cm}{4cm}{\includegraphics{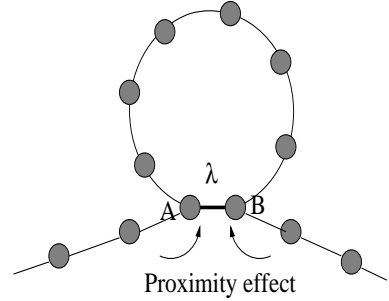}} \par}
\caption{A one dimensional ordered quantum wire is bent to bring 
two `end atoms' close enough to ensure a tunnel effect. 
$\lambda$ is the tunnel hopping integral.} 
\label{wire1}
\end{figure}

In this communication we present a blend of the spirits of the studies of
electronic transport in winding polymers and the model studies on QW systems 
within a single orbital tight binding scheme. 
Our interest is divide into two parts. First, We address the changes observed in the 
spectrum and transport characteristics of a QW modeled by an array of atomic sites, 
when the `wire' is bent so as to provide the traveling electron with an additional,  
{\it short cut} path to jump over to an otherwise distant site in the chain. 
The `closeness' between two sites in the chain, brought about by the `bending',  
will be modeled by an additional hopping amplitude along the shortcut path, and it's effect on the 
transmission spectrum will be discussed. 

Second, we examine the 
transmission spectrum of 
a quasi-periodic Fibonacci chain, bent and clamped between two semi-infinite ordered leads.
Usually, a Fibonacci quasi-periodic chain, in the thermodynamic limit, exhibits a completely
multifractal Cantor set spectrum~\cite{koh,nori,macia,macia2}. 
The fragmented character of the spectrum 
is reflected in a transport calculation of such chains clamped between the leads. We wish to 
investigate the effect of the short cut path generated as a result of bending the chain on the 
single particle transport in a Fibonacci chain of arbitrarily large size. 
This might throw light on the basic question of conductance of an aperiodic QW 
, or arrays od QD's when their geometry is deformed in some way so as to allow 
for any tunneling effect within the wire.

Our results are quite 
interesting. In the ordered lattice we find the creation of states localized at the points of 
bending and sharp asymmetric Fano like line shapes~\cite{fano,mirosh,arun} 
in the transmission, a simple case 
of which is worked out analytically here. The fragmented Cantor spectrum of a Fibonacci chain 
is found to be grossly affected by the bending, and all the gaps close with the onset of the 
hopping across the short cut path. The results, to the best of our knowledge, throw light on 
some of the basic facts that have not been discussed in details elsewhere.

In what follows we describe the results. In section I we present the model and the density of 
states of a bent ordered chain. Section II deals with the transmission spectrum of the 
bent ordered chain with an analysis of the Fano line shape in the simplest case. Section III 
deals with the quasi-periodic Fibonacci chain bent and clamped between two 
semi-infinite leads, and it's transport properties, and we draw conclusion in section IV.
\section{The Model and the energy spectrum: The periodic case}

We adopt a tight binding formalism, and incorporate only the nearest 
neighbor hopping. We begin by referring to Fig.~\ref{wire1}.  
We deal with non-interacting electrons and the single band Hamiltonian is given by, 
\begin{equation}
H=\sum_i \epsilon_i c_i^{\dagger} c_i + \sum_{<ij>} t_{ij}
\left(c_i^{\dagger} c_j +  c_j^{\dagger} c_i 
\right)
\label{equ1}
\end{equation}
where, 
$c_{i}$ ($c_{i}^{\dagger}$) are the annihilation (creation) operator at the 
$i$th site of the chain, $\epsilon_i$ is the on-site potential at the 
$i$-th site which we shall choose as $\epsilon_0$ for each site when the chain is
periodic. 
$t_{ij}$ is the nearest neighbor hopping integral, that will be taken as $t$ 
in the bulk portion of 
an ordered chain, and will be equal to $\lambda$ for hopping from the site 
$A$ to $B$ and vice versa. 
\begin{figure}[ht]
{\centering\resizebox*{18.3cm}{14.3cm}{\includegraphics [angle=-90] {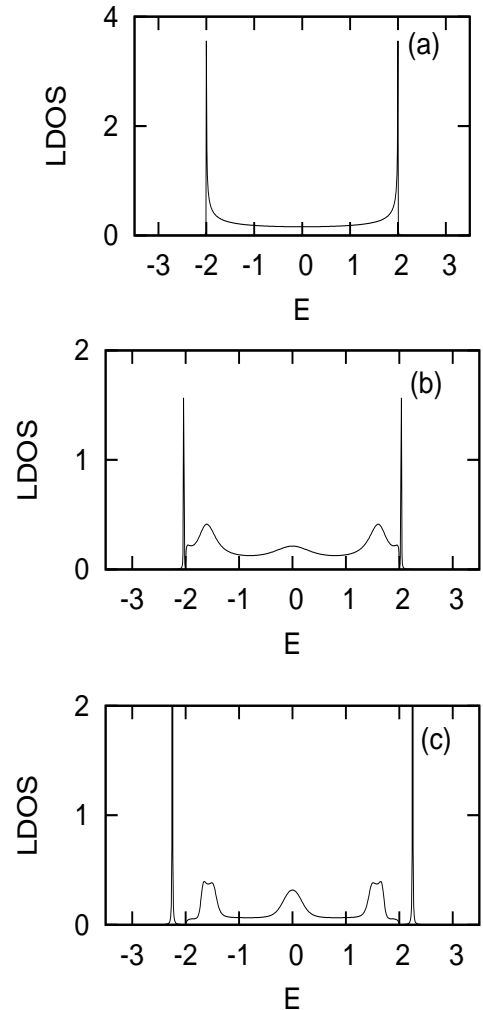}} \par}
\caption{Local density of states at the $A$ (or, $B$) site without and with bending. 
The figures correspond to  
(a) a purely linear chain with no bending ,i.e. $\lambda=0$ (dashed line), 
(b) $\lambda=0.6$  
and (c) $\lambda=1.0$ when the wire bends. Figures (b) and (c) refer to the 
situation when $N = 6$. We have chosen 
$\epsilon_0=0$ and $t=1$ and $E$ and $\lambda$ are measured in unit of $t$.}
\label{ldos}
\end{figure}

To obtain the local density of states (LDOS) at the junctions $A$ and $B$ 
of the bent system we 
make use of the system of equations satisfied by the Green's function 
$G_{ij}$, viz, 
\begin{equation}
(E-\epsilon_i) G_{ij} = \delta_{ij} +  \sum_k t_{ik} G_{kj}
\label{equ2}
\end{equation}
On the right hand side of Eq.~\ref{equ2} the index $k$ runs over the 
nearest neighbors of the $i$-th site. Obviously, $t_ik$ is equal to $t$ 
or, $\lambda$ as the case may be.  
To evaluate the LDOS at the site $A$ (or, $B$) we first re-normalize the portion of the 
chain `trapped' between the vertices $A$ and $B$ into an effective {\it diatomic molecule}
$AB$. The effective on-site potential at the site $A$  
is given by, 
\begin{equation}
\tilde \epsilon_A = \epsilon_0 + t \frac{M_{12}}{M_{11}}
\end{equation}
Naturally,
$\tilde \epsilon_B=\tilde \epsilon_A$. The re-normalize hopping integral 
connecting the $A$ site with the $B$ site (when the intermediate sites are 
decimated~\cite{southern}) is given by, 
\begin{equation}
t_{AB} = \lambda + \frac{t}{M_{11}}
\end{equation}
Here, $\lambda$ represents the short circuit hopping that results out of the 
bending of the wire, $M_{11} = U_N(x)$, and $M_{12} = -U_{N-1}(x)$ with $U_N(x)$ being 
the $N$th order Chebyshev polynomial of the second kind, and $x = (E-\epsilon_0)/2t$.
$E$ is the energy of the electron, and $N$ represents the number of atoms trapped in the 
bent portion of the chain in between the vertices $A$ and $B$.

The LDOS at the site $A$ (or, equivalently, at $B$) is then easily obtained 
by a well known decimation renormalization group (RG) method discussed 
elsewhere~\cite{southern}. In Fig.2 we show the LDOS at the site $A$ (or, $B$) when 
the chain is not folded (Fig.2a), and folded with increasing values of the 
short circuit hopping $\lambda$ (Fig.2b and 2c). 
While, in the first case, as expected, we get back the 
standard LDOS profile for a periodic chain, in the two latter cases the growth of two 
localized levels right beyond the values $E = \pm 2$ is apparent. As the short circuit 
hopping $\lambda$ increases, the localized levels move away from the band edges 
of the perfectly periodic open chain. The states are strictly localized at the edges $A$ 
and $B$. It is also interesting to note that apart from generating a couple of 
localized levels, a non-zero value of the cross hopping $\lambda$ changes the profile 
of the LDOS in a non-trivial fashion. In particular, the center of the band exhibits 
an oscillation in the values of the LDOS compared to the relatively flat low 
value in the case of an open chain,
as $\lambda$ increases. This feature can 
be taken to be a consequence of the proximity of the sites $A$ and $B$.

\section{Transmission characteristics}

We now address the transmission properties of a bent quantum wire 
modeled by the sites sitting in a periodic array and folded at the two 
points $A$ and $B$ as shown in Fig.1. To calculate the transmission coefficient 
we consider the folded portion of the infinite system as our `sample' and the 
remaining portions of the chain extending from the vertices $A$ and $B$ to infinity 
on either side as two semi-infinite ordered leads. 
The bent part of the system trapped between 
the vertices $A$ and $B$ is re-normalized~\cite{southern}, and we arrive at an effective 
diatomic molecule $A - B$ clamped between the leads. The effective on-site potentials
at the sites $A (B)$ are given by Eq.(3) and the effective hopping between 
the atoms $\tilde A$ and $\tilde B$ is given by $t_{AB}$ as we have already worked out. 
The transmission coefficient is given by the well known formula~\cite{stone}, 
\begin{widetext}
\begin{equation}
T = \frac{4 \sin^2ka}{[M_{12}-M_{21} + (M_{11}+M_{22})\cos ka]^2 
+ (M_{11}+M_{22})^2 \sin^2ka}
\end{equation}
\end{widetext}
where, $M_{11} = U_N(x)$, and $M_{12} = -M_{21} = -U_{N-1}(x)$, 
and $M_{22} = -U_{N-2}(x)$. $U_N(x)$ is 
the $N$th order Chebyshev polynomial of the second kind, $x = (E-\epsilon_0)/2t$, 
and $ka = \cos^{-1}[(E-\epsilon_0)/2t]$. The lattice spacing $a$ of the chain is 
taken to be unity in all results presented here. Before presenting any general result we 
draw the attention of the reader to a special feature of the transmission profile, viz, the 
appearance of the Fano line shape~\cite{fano,arun}. 
\begin{figure}[ht]
{\centering\resizebox*{7.3cm}{4.8cm}{\includegraphics{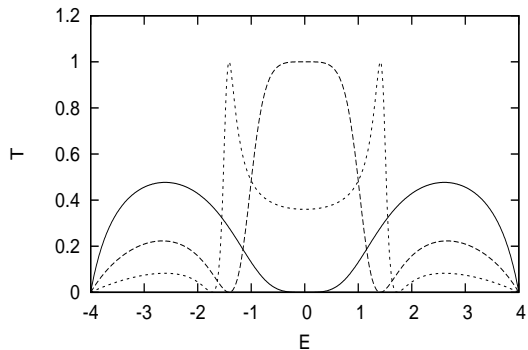}} \par}
\caption{Transmission across a bent ordered chain with $N=2$ when 
$\lambda=2$ (solid line), $\lambda=4$ (dashed line) and $\lambda=8$ (dotted
line). 
We have taken $\epsilon_0=0$ and $t=1$, and $\lambda$ is measured in 
unit of $t$.}
\label{trans1}
\end{figure}
\begin{figure}[ht]
\begin{center}
\includegraphics[height=5cm,width=7cm,angle=0]{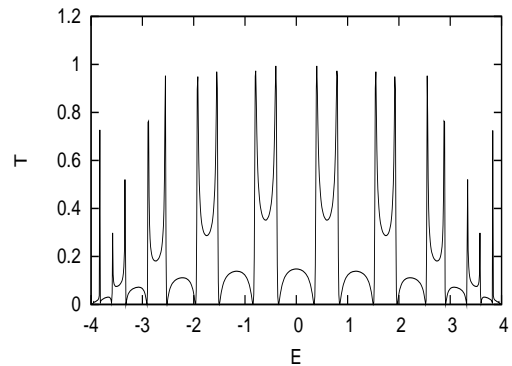}
\caption{ 
 Transmission across an bent ordered chain of $22$ atoms with $\lambda=8$.
Asymmetric Fano line shape appears at every anti-resonance point.
Other parameters are the same as in Fig.4}
\label{trans2}
\end{center}
\end{figure}

To get a clear understanding, we work out the very simple 
case of bending when there is just one atom in the arc between the junctions $A$, and $B$.
That is, we take $N=1$.
In this case, the on-site potentials at the end atoms, and the effective hopping $t_{AB}$ 
of the `diatomic molecule' clamped between the leads are given by, 
\begin{eqnarray}
\tilde\epsilon_A & = & \epsilon_0 + \frac{t^2}{E-\epsilon_0} \nonumber \\
t_{AB} & = & \lambda + \frac{t^2}{E-\epsilon_0} 
\end{eqnarray}
Clearly, an anti-resonance will occur at 
\begin{equation} 
E = \epsilon_0 - \frac{t^2}{\lambda}
\end{equation}
Evidently, as $\lambda \rightarrow \infty$ (atomic sites $A$ and $B$ coming 
indefinitely close together), the anti-resonance shifts towards the center of the 
spectrum, i.e. towards $E=\epsilon_0$. 
It is quite straightforward (at the expense of a little algebra though) 
to show that as $\lambda \rightarrow \infty$, then in a $\delta$-neighborhood of the 
anti-resonance , i.e., for $E = \epsilon - t^2/\lambda + \delta$, such that the product 
$\lambda\delta$ remains finite, the transmission amplitude $\tau$ ($T = |\tau|^2$) 
can be approximately written as, 
\begin{equation}
\tau = e^{i\pi}\frac{4t}{\lambda} \left [\frac{\delta}{(\delta+\frac{t^2}{\lambda}) - 
i \frac{2t^3}{\lambda^2}} \right ]
\end{equation}
The term within the square bracket controls the line shape of the transmission 
characteristic around the point of anti-resonance while the pre-factor $4t/\lambda$ gives 
the amplitude. Two points are to be noted. First, as $\lambda$ increases, the amplitude 
of transmission drops. Second, The numerator inside the square bracket is simply $\delta$, 
which will be zero as $\delta \rightarrow 0$, while the real part of the denominator 
in the square bracketed term has a zero at $\delta = -t^2/\lambda$. These de-tuned zeros give 
rise to an asymmetric Fano line shape around the point of anti-resonance as is 
already discussed 
in the literature~\cite{arun,voo}. The width of the transmission profile is of course, given 
by the remaining term $2t^3/\lambda^2$.

For $N > 1$, the analytical attack to reveal any Fano profile, though not impossible, 
becomes a bit complicated, and a straightforward use of Eq. (5) is advised for the numerical 
evaluation of the transmission coefficient. Fig.3 displays the features for $N = 2$ with 
$\lambda = 2t$ (solid line), $4t$ (dashed line) and $8t$ (dotted line) respectively.
For $\lambda = 2t$, the transmission is low in general (compared to an open periodic chain, 
and hinting to the fact that the proximity effect induces some destructive interference), 
with a broad minimum around the center of the spectrum. As $\lambda$ grows the spectrum 
is marked by the appearance of a couple of transmission resonance peaks for $\lambda=4t$,
which merge into a broad maximum when $\lambda = 8t$. 
The development of 
asymmetric Fano profile around the $T=0$ points is obviously on the cards, and becomes 
much more prominent as we increase $N$, the number of atoms trapped in the bent portion (Fig.4).

\section{Transmission in a bent Fibonacci wire}

We now present results for a bent quasi-periodic Fibonacci chain 
clamped between two semi-infinite leads in such a manner that the 
lead-sample connecting points are close enough to ensure a tunnel-hopping 
across the junctions (Fig.5). 
\begin{figure}[ht]
\begin{center}
\includegraphics[height=5cm,width=4cm,angle=0]{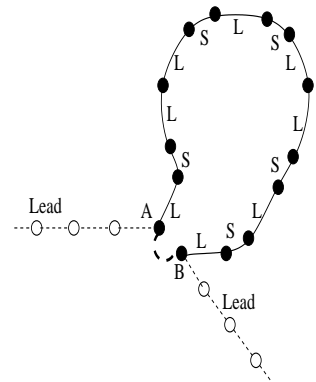}
\caption{A quasi-periodic Fibonacci chain (transfer model) is bent to bring 
the `end atoms' in close proximity. The {\it short cut} hopping is shown 
as the dotted line connecting $A$ and $B$.} 
\label{wire2}
\end{center}
\end{figure}

The Fibonacci chain is the canonical example 
of quasi-periodic order in one dimension. The eigenvalue spectrum is known to be 
a Cantor set with measure zero~\cite{koh}. The chain is grown recursively using two 
symbols $L$ and $S$ that stand for bonds with two different hopping strengths, viz, 
$t_L$ and $t_S$ in our case. The growth rule is, $L \rightarrow LS$ and $S \rightarrow L$, 
with the first generation chain comprising of a single $L$-bond. 
In the corresponding tight binding Hamiltonian, the on-site potentials assume 
three possible values depending on the nearest neighbor configuration of any vertex, 
viz, $\epsilon_\alpha$, $\epsilon_\beta$ and $\epsilon_\gamma$ when an atomic site 
is flanked by the bonds $L-L$, $L-S$ and $S-L$ on its two sides respectively.

Without losing any generality, 
we consider odd generation chains clamped between the ordered leads. The trapped portion 
of the chain is then re-normalized using the recursion relations, 
\begin{eqnarray}
\epsilon_\alpha(n+1) & = & \epsilon_\alpha(n) + 
\frac{t_L(n)^2 (2E-\epsilon_\beta(n)-\epsilon_\gamma(n))}{\delta(n)} \nonumber \\
\epsilon_\beta(n+1) & = & \epsilon_\alpha(n) + 
\frac{t_L(n)^2 (E-\epsilon_\beta(n))}{\delta} + 
\frac{t_L(n)^2}{E-\epsilon_\beta(n)} \nonumber \\
\epsilon_\gamma(n+1) & = & \epsilon_\gamma(n) + 
\frac{t_L(n)^2 (E-\epsilon_\gamma(n))}{\delta} + 
\frac{t_S(n)^2}{E-\epsilon_\beta(n)} \nonumber \\
t_L(n+1) & = & \frac{t_L(n)^2 t_S(n)}{\delta(n)} \nonumber \\
t_S(n+1) & = & \frac{t_L(n)t_S(n)}{E-\epsilon_\beta(n)}
\end{eqnarray}
where, $\delta(n) = (E-\epsilon_\beta(n))(E-\epsilon_\gamma(n))-t_S(n)^2$.
\vskip .25in
As in the previous section, the transmission coefficient is obtained after decimating 
~\cite{arun} all the sites of the Fibonacci chain trapped in the bent portion thereby creating an 
effective diatomic molecule as before. Fig. 6 shows the transmission spectrum for the set of values 
$\epsilon_\alpha = \epsilon_\beta = \epsilon_\gamma =0$, $t_L = 1 = t_S/2$ and 
$\lambda = 8$. It is interesting to observe that all the gaps which are present in 
an open Fibonacci chain are absent in the presence of the short cut hopping $\lambda$. 
\begin{figure}[ht]
{\centering\resizebox*{12.5cm}{10cm}{\includegraphics{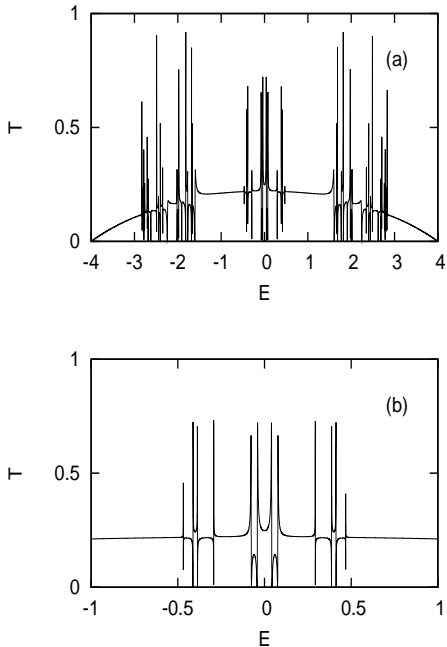}} \par}
\caption{Transmission across a $9$-th generation 
Fibonacci chain with proximity effect. $\lambda=8$, 
$\epsilon_\alpha=\epsilon_\beta=\epsilon_\gamma=0$, 
$t_L=1$ and $t_S=2$.} 
\label{trans3}
\end{figure}

Asymmetric Fano line shape is now found to be associated with 
every anti-resonance in the transmission spectrum. Fig.6b displays the magnified 
version of the $T-E$ graph between $E = \pm 1$ ($E$ is measured in unit of $t_L$ which we 
select as unity). It is interesting to observe that the figure indeed displays 
the characteristic self-similarity of the spectrum of a quasi-periodic open chain, but now 
every gap that is found in the spectrum of an open Fibonacci chain is found to be closed. This 
is the effect of the proximity of the two extreme points $A$ and $B$ in the lattice. The 
Fano line shape becomes much more prominent.

Though, we have shown in Fig.6 the transmission spectrum for a somewhat bigger value of
the cross hopping $\lambda$, extensive numerical investigation reveals that the closing of the 
gaps, at every scale of the energy, happens whenever the tunneling of the electron is 
allowed between the contact points directly, i.e., even with a very small value of $\lambda$. 
This opens up a question regarding the stability of the cantor set eigenvalue spectrum in a 
realistic Fibonacci quantum wire that is bent so as to bring it's two ends to close 
proximity.

\section{Conclusion}
In conclusion, we have addressed the question of two terminal transmission across 
a linear chain when a portion of it is bent so as to bring two distant atomic sites 
into close proximity. An ordered chain and a quasi-periodic chain are discussed. For
the ordered case it is shown 
that, the bending and the corresponding cross hopping of electrons generates Fano line shape
in the transmission profile, a simple example of which 
is analytically examined.  
The overall transmittance is reduced as the cross hopping is switched on, and some 
resonance peaks mark the spectrum.
Localized states are created with a non-trivial change 
in the central part of the energy spectrum. For the quasi-periodic chain, 
the bending destroys the 
gaps in the spectrum at all scales of energy. The self-similar character is somewhat 
retained, and Fano like line shapes are likely to mark every transmission anti-resonance.
On finer scan of the energy interval, the closing of the gaps becomes more and more 
apparent.
\vskip .3in
\begin{center}
{\bf ACKNOWLEDGMENT}
\end{center}
\vskip .25in
Illuminating conversations with Rudolf A. R\"{o}mer, Alberto Rodriguez, and 
Santanu Maiti are gratefully acknowledged.

\end{document}